\documentclass[conference]{IEEEtran}
\IEEEoverridecommandlockouts

\newcommand\hmmax{0}
\newcommand\bmmax{0}

\usepackage{graphicx}
\usepackage{tikz}
\usepackage{pgfplots}
\usepackage{xcolor}
\usepackage{color, colortbl}
\usepackage{amsmath}
\usepackage{amsfonts, amssymb}
\usepackage{mathtools}
\usepackage{cite}
\usepackage[nolist]{acronym}
\usepackage{textcomp}
\usepackage{upgreek}
\usepackage{siunitx}
\usepackage{booktabs} 		
\usepackage{diagbox}		
\usepackage{upgreek}
\usepackage{bbm}
\usepackage{Files/bm}
\usepackage{Files/winsnotation}
\usepackage{Files/Symbols_OPL}
\usepackage{Files/LVcolours} 

\usepackage[ruled,vlined,noresetcount]{algorithm2e}
\usepackage{setspace}	 	

\usepackage{array}
\newcolumntype{C}[1]{>{\centering\arraybackslash}p{#1}}

\makeatletter
\def\endthebibliography{%
  \def\@noitemerr{\@latex@warning{Empty `thebibliography' environment}}%
  \endlist
}
\makeatother
\usepackage{amsthm}
\theoremstyle{definition}
\newtheorem*{remark}{Remark}

\newcommand{\dB}{\mathrm{dB}}
\newcommand{\rmt}t
\newcommand{\rmr}r

\pgfplotsset{compat=1.17}

\begin{document}

\title{Analysis of Pointing Loss Effects \\ in Deep Space Optical Links}

\author{%
  \IEEEauthorblockN{Lorenzo Valentini, Alberto Faedi, Enrico Paolini, Marco Chiani}
  \IEEEauthorblockA{CNIT, DEI, University of Bologna,
  	Cesena, Italy\\
  	Email: \{lorenzo.valentini13, e.paolini, marco.chiani\}@unibo.it }
}

\maketitle 

\begin{acronym}
\small
\acro{AU}{astronomic unit}
\acro{CDF}{cumulative distribution function}
\acro{CCSDS}{Consultative Committee for Space Data Systems}
\acro{CRC}{cyclic redundancy check}
\acro{EDRS}{European Data Relay Satellite}
\acro{ESA}{European Space Agency}
\acro{FER}{frame error rate}
\acro{GEO}{geosynchronous equatorial orbit}
\acro{i.i.d.}{independent and identically distributed}
\acro{ISS}{International Space Station}
\acro{ITU}{International Telecommunication Union}
\acro{LEO}{low-Earth orbit}
\acro{LLCD}{Lunar Laser Communications Demonstration}
\acro{LOS}{line of sight}
\acro{MLCD}{Mars Laser Communications Demonstration}
\acro{NASA}{National Aeronautics and Space Administration}
\acro{OPALS}{Optical PAyload for Lasercomm Science}
\acro{PDF}{probability density function}
\acro{PMF}{probability mass function}
\acro{PPM}{pulse-position modulation}
\acro{RF}{radio-frequency}
\acro{SCPPM}{serially concatenated pulse-position modulation}
\acro{SILEX}{Semiconductor-laser Intersatellite Link EXperiment}
\acro{TM}{telemetry}
\end{acronym}
\setcounter{page}{1}

\begin{abstract}
Owing to the extremely narrow beams, a main issue in optical deep space communications is represented by miss-pointing errors, which may severely degrade the system performance and availability.   
In this paper, we address pointing losses in the case in which both the receiver and the transmitter are affected by angular errors.  
Pointing losses are evaluated through two approaches. The first approach is deterministic and only requires knowledge of a maximum angular error. 
The second approach requires knowledge of the angular error statistical distribution and tackles the problem from an outage probability viewpoint. 
These tools are then applied to analyze the impact of pointing losses in deep space optical links in which both terminals suffer from miss-pointing effects. The antenna gains are first optimized to maximize the effective system gain. The optimum antenna gains are then applied to evaluate maximum achievable ranges and to perform link design by means of optical link budgets. 
\end{abstract}


\section{Introduction}

Free space optical links offer an attractive alternative in wireless communications, for data transmission at very high rates over long distances. 
Licence-free spectrum and low mass and power requirements are features fostering the use of optical frequencies in several application domains \cite{HemBisDjo:11,Kau:17}, including high-speed transportation, unmanned aerial vehicles, building-to-building communications, satellite and deep space communications~\cite{Cha:06}. 
The raising demand for \ac{TM} data from space and the growing volumes of such data, in particular, are contributing to the increasing interest for optical bands in space communications, with the expectation to boost data rates by two orders of magnitude over conventional \ac{RF} links \cite{Lay:97, HemBisDjo:11}.

In the recent past, several space missions have exploited optical frequencies.
In 2001, the \ac{ESA} implemented and tested an inter-satellite laser link within the \ac{SILEX} mission.
In 2009, the \ac{NASA} \ac{MLCD} proved the feasibility of deep space optical communications; moreover, in 2013, the \ac{NASA} \ac{LLCD} mission demonstrated high-rate laser communications from a small terminal at lunar distances.
In 2014, the \ac{NASA} \ac{OPALS} mission aimed at demonstrating optical communications between the \ac{ISS} and Earth ground stations. 
In 2016 and 2019, the \ac{ESA} \ac{EDRS} mission sent into \acl{GEO} two satellites with the goal of providing near global coverage for satellites in \acl{LEO}, exploiting optical links. 

Although the above-mentioned missions have confirmed that optical frequencies can achieve higher data rates with lower size and mass with respect to the \ac{RF} case, their use in space links also poses several challenges. 
In this respect, main issues are represented by pointing errors, solar conjunctions, and atmospheric effects \cite{Vil:81,Zhu:02,Far:07, XuZen:19}.
Even in an inter-satellite scenario, where atmospheric effects are absent, the pointing losses alone can frustrate the benefits offered by the optical frequencies. 
Hence, several studies have been conducted to model and analyze these losses \cite{LyrPan:19, BisHemPia:10, BisPia:03, HemBisDjo:11, ITU:06}. 
The extremely narrow optical beams, essential in deep space communications to maximize the received power, pose very tight requirements in terms of pointing accuracy. 
As a consequence, the miss-pointing losses between the transmit and the receive antennas cannot be neglected even in presence of a very accurate point-ahead calculation, due to mechanical noise generated by the vibrations and the satellite motions \cite{CheGar:89}. 

Most available studies on pointing effects assume that only one antenna is affected by angular errors \cite{Vil:81}. 
For example, when one edge of the communication link is on ground, a very precise ground pointing can be assumed and, therefore, it is usual to consider the angular noise only on the spacecraft.
In case of space-to-space links, however, pointing effects should be carefully considered at both edges of the link. 
A particularly interesting situation where this problem arises is represented by ``two-hop'' deep space communication architectures, featuring a first optical link between a deep space probe and a relay orbiting the Earth (to avoid atmospheric impairments), and a second link between the relay and the Earth using, e.g., a classical \ac{RF} frequency. 
Therefore, in this paper we investigate approaches to address the miss-pointing losses in situations where both terminals are affected by angular noise. 
The concept of \emph{effective system gain} is introduced and the effect of pointing losses is investigated in terms of maximum achievable ranges and of deep space optical link budgets. 

This paper is organized as follows. Section~\ref{sec:preliminary} addresses angular error distributions and pointing loss models. 
Section~\ref{sec:Approaches} describes two approaches to pointing losses, the first based on a maximum angular error and the second on the angular error statistic. 
Section~\ref{sec:NumericalResults} addresses gain optimization, with application to maximum achievable ranges and optical link budgeting. Conclusions are drawn in Section~\ref{sec:conclusions}.

\section{System Model}
\label{sec:preliminary}

In this section we introduce our assumptions as well as the pointing error descriptions over free space communication links. 
We also report three pointing loss models that can be used according to the antenna features \cite{CheGar:89, BisPia:03}. 
The combination of these descriptions allows analyzing and addressing the communication limits due to pointing errors.

\subsection{Angular Error Distribution}
Pointing in a three-dimensional space is affected by both azimuth and elevation angular errors. 
In this paper we assume that these two errors are \ac{i.i.d.} and Gaussian distributed with zero-mean. 
Therefore, the resulting random angular error $\theta$, the angle between the \ac{LOS} and the pointing direction, is Rayleigh distributed as \cite{CheGar:89, BarMec:85}
\begin{align}
\label{eq:Rayleigh}
f_{\theta}(\theta) = \frac{\theta}{\sigma_{\theta}^{2}} \, e^{-\theta^2/2\sigma_{\theta}^{2}} , \quad \theta \geq 0
\end{align}
where $\sigma_{\theta}$ is the main angular noise parameter, hereafter referred to as the \emph{pointing accuracy}. 
Another model, sometimes used in the literature, assumes the azimuth and elevation angular errors as independent and Gaussian, but with a non-zero mean. This yields an angular error that is Rician-distributed according to
\cite{Vil:81,Kau:17}
\begin{align}
    f_{\theta}(\theta) = \frac{\theta}{\sigma_{\theta}^{2}} \, e^{-(\theta^2 + \eta^2)/2\sigma_{\theta}^{2}} \, I_0\left(\frac{\theta\,\eta}{\sigma_\theta^2}\right) , \quad \theta \geq 0
\end{align}
where $I_0(\cdot)$ is the modified Bessel function of the first kind of order $0$, and $\eta$ is the bias error angle from the center.

\subsection{Pointing Loss Models}
\begin{figure}[t]
    \centering
    \resizebox{0.95\columnwidth}{!}{
    	\input{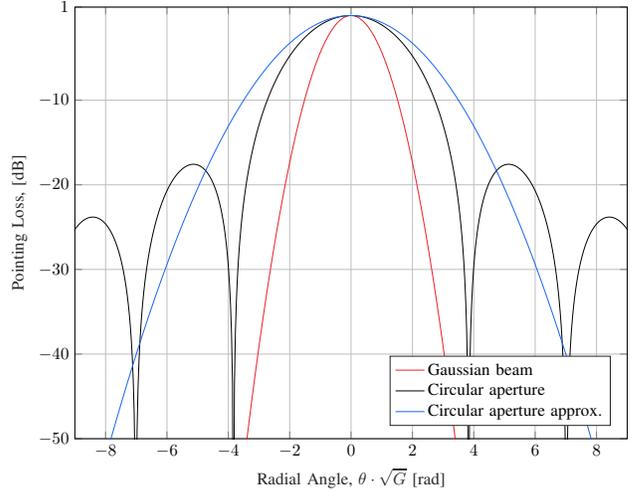}
    }
    \caption{Comparison between different pointing loss models.}
    \label{fig:PointingLossModels}
\end{figure}

The pointing losses are described by the radiation pattern of the antenna. 
In fact, they can be seen as a ``correction factor'' accounting for the gain $G$ being set equal to its maximum value $G_\mathrm{max}$. 
Hence, in order to analyze these losses, a description of the antenna gain, parameterized by the angular miss-pointing along the \ac{LOS} direction, is required. 
In the following we summarize the models used in this paper, where $G$ is the linear-domain antenna gain and $\theta$ is the angular miss-pointing error. 
In case of Gaussian beam, the pointing losses are given by \cite{CheGar:89}
\begin{align}
    L_\mathrm{p}(\theta) = e^{-G\theta^2}\,.
    \label{eq:GaussianBeamLosses}
\end{align}
In case of circular aperture, the pointing losses are instead given by \cite{CheGar:89,Orf:04}
\begin{align}
\label{eq:CircModel}
    L_\mathrm{p}(\theta) = \left( \frac{2 \, J_1(\sqrt{G}\theta)}{\sqrt{G}\theta} \right)^2
\end{align}
where $J_1(\cdot)$ is the Bessel function of the first kind of order 1. 
This latter model can be well-approximated, for small angular error $\theta$, as \cite{Mar:20}
\begin{align}
\label{eq:approxCirc}
    L_\mathrm{p}(\theta) = \left( \frac{2 \, J_1(\sqrt{G}\theta)}{\sqrt{G}\theta} \right)^2 \approx e^{-\alpha G\theta^2}
\end{align}
where $\alpha = 0.188$. Fig.~\ref{fig:PointingLossModels} illustrates the pointing losses computed according to the above models.
%

\section{Pointing Losses} 
\label{sec:Approaches}

In this section we address two approaches to quantify pointing losses. 
The first approach treats the angular error as a \emph{deterministic parameter} (usually regarded as the maximum angular error); the second one considers the angular error as a \emph{random variable}, which allows formulating the problem in terms of an outage probability. 

\subsection{Deterministic Approach}
The deterministic approach consists of simply taking a maximum angular error $\theta_\mathrm{max}$ (based on the available information about the tracking system) and of quantifying the pointing loss by employing one of the above-presented models, using the numerical value of $\theta_\mathrm{max}$. 
This  method is useful in case no a priori knowledge is available about the statistical distribution of the angular error. 
For example, using a Gaussian beam model, for given $\theta_\mathrm{max}$ we compute the maximum pointing loss to be included in the power budget as
\begin{align}
L_\mathrm{p, max} = e^{-G \theta^2_\mathrm{max}} .
\end{align}
This procedure is simple but provides no information about the probability that the actual pointing loss $L_{\mathrm p}$ exceeds the value $L_\mathrm{p, max}$ during its random fluctuations. 
In other words, it is unable to capture the concept of outage.

\subsection{Outage Probability Approach}
In contrast with the deterministic approach, we propose a statistical approach introducing a pointing outage probability defined as
\begin{align}
P_\mathrm{out} &= \Prob{A_\mathrm{p}[\dB] > A_\mathrm{p}^{*} [\dB]}
\end{align}
where $A_{\mathrm p}[\dB]=-L_\mathrm{p}[\dB]$ and $A_\mathrm{p}^{*}[\dB]$ is the margin against the pointing attenuation to be included in the link budget. 
Hence, we define the outage probability as the probability that the pointing loss exceeds the maximum value we can compensate with the employed margin $A_\mathrm{p}^{*}[\dB]$.

In case of a link between two antennas, both affected by mechanical noise, it is necessary to consider the pointing losses introduced by each of them. In this situation, we can write
\begin{align}
P_\mathrm{out} = \Prob{A_{\mathrm{p},\rmt}[\dB] + A_{\mathrm{p},\rmr}[\dB] > A_\mathrm{p}^{*} [\dB]}
\end{align}
where the subscripts `$\rmt$' and `$\rmr$' refer to the transmitter and the receiver, respectively. Hence, to compute the outage probability both error distributions and antenna gain radiation pattern are required. 

In case both radiation patterns follow the Gaussian beam model and the two angular errors are independent and Rayleigh distributed with parameters $\sigma_{\theta,t}$ and $\sigma_{\theta,r}$, we obtain
\begin{align}
    P_\mathrm{out} = \Prob{G_{t} \theta^2_{t} + G_{r}\theta^2_{r} > K}
\end{align}
where $K = \frac{\ln(10)}{10}\,A_\mathrm{p}^{*}[\dB]$. 
Substituting $y_r = \theta^2_r$ and $y_t = \theta^2_t$, we obtain two independent exponential random variables with mean values $2\sigma_{\theta,t}^2$ and $2\sigma_{\theta,r}^2$, yielding
\begin{align}
\label{eq:GauBeamPout}
    P_\mathrm{out} 
    &= 1 - \! \int_{0}^{K/G_\rmr} \!\!\! \int_{0}^{\frac{K-G_\rmr y_\rmr}{G_\rmt}} \!\!\!\! \frac{1}{4\sigma_{\theta,\rmt}^2 \sigma_{\theta,\rmr}^2}\,
    e^{-\left[ \frac{y_\rmt}{2 \sigma^2_{\theta,\rmt}} + \frac{y_\rmr}{2 \sigma^2_{\theta,\rmr}} \right]}
    \mathrm{d}y_\rmt \mathrm{d}y_\rmr \nonumber \\
    &= \frac{G_\rmt\sigma_{\theta,\rmt}^{2} e^{-K /(2\sigma_{\theta,\rmt}^{2}G_\rmt)}}{G_\rmt\sigma_{\theta,\rmt}^{2}-G_\rmr\sigma_{\theta,\rmr}^{2}}\, + \frac{G_\rmr\sigma_{\theta,\rmr}^{2} e^{-K /(2\sigma_{\theta,\rmr}^{2}G_\rmr)}}{G_\rmr\sigma_{\theta,\rmr}^{2}-G_\rmt\sigma_{\theta,\rmt}^{2}}\,.
\end{align}
For given $P_{\mathrm{out}}$, antenna gains, and pointing accuracy parameters, \eqref{eq:GauBeamPout} allows obtaining $A_\mathrm{p}^{*}$ (through $K$) numerically. In the particular case $G_\rmt = G_\rmr = G$ and $\sigma_{\theta, \rmt} = \sigma_{\theta, \rmr} = \sigma_{\theta}$, the outage probability becomes
\begin{align}
\label{eq:GauBeamPoutGEqual}
P_\mathrm{out} = \left(1 + \frac{K}{2\sigma_{\theta}^{2}G}\right)\,e^{-K/2\sigma_{\theta}^{2}G}\,.
\end{align}

Very similar results are obtained, by introducing the parameter $\alpha$, for the circular aperture model through the approximation described in \eqref{eq:approxCirc}. 
This approximation is tight when pointing losses accounted for in the link budget $A_\mathrm{p}^{*}$ are small (small angular error) as depicted in Fig.~\ref{fig:PointingLossModels}. 
We simply have
\begin{align}
    P_\mathrm{out} =
    \frac{G_\rmt\sigma_{\theta,\rmt}^{2} e^{-K / (2\sigma_{\theta,\rmt}^{2}\alpha G_\rmt)}}{G_\rmt\sigma_{\theta,\rmt}^{2}-G_\rmr\sigma_{\theta,\rmr}^{2}}\, + \frac{G_\rmr\sigma_{\theta,\rmr}^{2} e^{-K /(2\sigma_{\theta,\rmr}^{2}\alpha G_\rmr) }}{G_\rmr\sigma_{\theta,\rmr}^{2}-G_\rmt\sigma_{\theta,\rmt}^{2}}\,.
\end{align}

Different models or error distributions may require numerical evaluation. 
Also note that, in case only one antenna (either on the transmit or on the receive side) is affected by miss-pointing errors, we can ``artificially'' obtain the outage probability by simply letting either $G_\rmr$ or $G_\rmt$ be null. 
For example, under a Gaussian beam model and letting miss-pointing error affect the receiver only, setting $G_\rmt=0$ and $G_\rmr=G$ in \eqref{eq:GauBeamPout}, we obtain
\begin{align}
\label{eq:GauBeamPoutSingleG}
P_\mathrm{out} = e^{-K/2\sigma_{\theta}^{2}G}\,.
\end{align}
This is useful when one of the two losses dominates the other or when the effect of pointing errors on one side is negligible.


\section{Application to Deep Space Optical Links}
\label{sec:NumericalResults}
Hereafter we focus on a deep space optical \ac{TM} link with photon counting receiver and Poisson channel model \cite{VilBis:02, MoiHam:03}. 
The considered coding and modulation scheme is the \ac{SCPPM} one, recommended by the \ac{CCSDS} for the Coding and Synchronization Sublayer of deep space optical \ac{TM} \cite{MoiHam:05,CCSDS:19}. 
The encoder features a serial concatenation of a convolutional encoder, an interleaver, an accumulator, and a pulse position modulator. 
The advantage of \ac{SCPPM} lays on the possibility to iteratively decode the received symbols achieving near-capacity performance \cite{MoiHam:03}.\footnote{We point out the existence of other competitive coding schemes for \ac{PPM} optical communications, such as binary LDPC codes \cite{Tan2008:coded_modulation,Zhou2013:protograph} and nonbinary LDPC codes \cite{Matuz2017:NBLDPC}.}
The \ac{PPM} is often employed in deep space links, where peak and average power constraints are typical \cite{Lip:80, MoiHam:05b, SimVil:04}. 
The output of the \ac{PPM} modulator is a symbol composed of $M=2^m$ slots, of which only one is pulsed; the position of the laser pulse in a symbol is given by the configuration of the corresponding $m=\log_2 M$ input bits. 
Over a Poisson channel, the receiver observes a number $k$ of photons in each slot distributed according to
\begin{align}
    \begin{split}
        &\Prob{k \, | \, \text{slot pulsed}} = \frac{\left( n_\mathrm{s} + n_\mathrm{b} \right)^{k}}{k!} \, e^{-\left( n_\mathrm{s} + n_\mathrm{b} \right)}\\
        &\Prob{k \, | \, \text{slot not pulsed}} = \frac{n_\mathrm{b} ^{k}}{k!} \, e^{-n_\mathrm{b}}
        \label{eq:PMFpoissonProcess}
    \end{split}
\end{align}
where $n_\mathrm{s}$ is the mean number of signal photons per slot, and $n_\mathrm{b}$ is the mean number of noise photons per slot.

\subsection{Antenna Gain Optimization}
\label{sec:Optimization}
Regardless of the approach used to estimate the pointing losses and the adopted model, we can define an ``effective system gain'' as
\begin{align}
G_\mathrm{eff} = (G_\rmt L_{\mathrm{p},\rmt} ) \, (G_\rmr L_{\mathrm{p},\rmr} )
\end{align}
i.e., as the product of the transmit and receive antenna gains, each multiplied by the corresponding pointing loss. 
Looking at the effective system gain reveals the existence of an optimal value for the antenna gains, beyond which the overall system performance degrades instead of improving. 
Considering a situation where the transmitter and the receiver are equipped with the same antennas, the effective system gain, expressed in dB, assumes the form
\begin{align}
G_\mathrm{eff}\,[\dB] = 2G \,[\dB] - A_\mathrm{p,tot} \,[\dB]
\end{align}
where $A_\mathrm{p,tot}[\dB]$ incorporates the total pointing losses. 
Sticking for simplicity to the case in which both antennas exhibit the same pointing accuracy parameter, defined either by $\sigma_\theta$ or by $\theta_\mathrm{max}$ depending on the employed approach, we can numerically find the total pointing attenuation $A_\mathrm{p,tot}[\dB]$ (and therefore the numerical value of $G_\mathrm{eff}[\dB]$) for any given gain $G[\dB]$. 
\begin{figure}[t]
    \centering
    \resizebox{\columnwidth}{!}{
    	\input{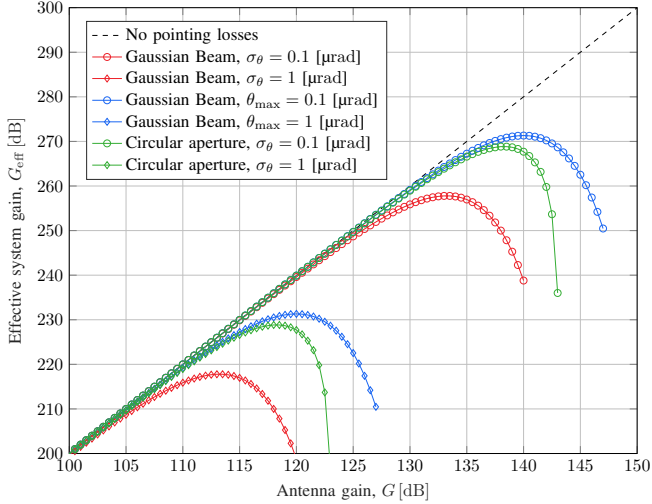}
    }
    \caption{Effective system gain versus the antenna gain. Deterministic and probabilistic approach to the pointing losses; different values of the pointing accuracy parameters; Gaussian beam and circular aperture models.}
    \label{fig:EffectiveGain}
\end{figure}

An example is shown in Fig.~\ref{fig:EffectiveGain} for different values of the pointing accuracy parameter ($\sigma_\theta$ or $\theta_\mathrm{max}$) and different pointing loss models (Gaussian beam or circular aperture). 
In case the probabilistic approach is used, the outage probability is set to $P_\mathrm{out} = 5\%$. 
The red and green curves correspond to the probabilistic approach for the Gaussian beam and circular aperture models, respectively; the blue curves refer to the deterministic approach in the case of Gaussian beam.
\begin{remark}
The optimal gain in the Gaussian beam case, when the approach based on $\theta_\mathrm{max}$ is used, can be easily derived as 
\begin{align}
    G_\mathrm{opt} = \frac{1}{\theta_\mathrm{max}^2}\,.
    \label{eq:GoptGaussianTheta}
\end{align}
\end{remark}

%
\begin{table*}[t]
\centering
\caption{Maximum achievable ranges, in AU, with a given PPM order and a specified $\sigma_{\theta}$. Results found using Gaussian beam model, $R = 1/3$, $T_\mathrm{s} = 256 \, \mathrm{ns}$, $n_\mathrm{b} = 1.21\cdot 10^7\, \mathrm{phe/s}$, $P_{\mathrm{av}} = 5\,\mathrm{W}$, $P_\mathrm{out}=5$\%, $\lambda = 1064\,\mathrm{nm}$, Link Margin $3\,\dB$.}
\footnotesize
\begin{tabular}{l|C{1.6cm}C{1.6cm}C{1.6cm}C{1.6cm}C{1.6cm}C{1.6cm}C{1.6cm}}
\toprule
 & \multicolumn{7}{c}{\textbf{PPM Order / Peak Laser Power [W]}} \\
 $\sigma_{\theta} [\mathrm{\upmu rad}]$ & 256 / 1600 & 128 / 800 & 64 / 400 & 32 / 200 & 16 / 100 & 8 / 50 & 4 / 25 \\
\midrule
1.00 & 0.113 & 0.084 & 0.062 & 0.046 & 0.034 & 0.025 & 0.019\\
0.50 & 0.453 & 0.336 & 0.249 & 0.184 & 0.136 & 0.101 & 0.075\\
0.35 & 0.924 & 0.684 & 0.507 & 0.375 & 0.277 & 0.206 & 0.152\\
0.20 & 2.836 & 2.100 & 1.555 & 1.151 & 0.851 & 0.631 & 0.467\\
0.15 & 5.037 & 3.730 & 2.762 & 2.045 & 1.512 & 1.121 & 0.830\\
0.10 & 11.342 & 8.398 & 6.218 & 4.605 & 3.406 & 2.525 & 1.869\\
0.05 & 45.350 & 33.580 & 24.864 & 18.411 & 13.617 & 10.094 & 7.474\\
\midrule
Data Rate [$\mathrm{kbps}$] & 32.33 & 56.58 & 97.00 & 161.66 & 258.66 & 387.99 & 517.32\\
\bottomrule
\end{tabular}
\label{tab:MaxDistSigma}
\end{table*}
%
\begin{table*}[t]
\centering
\caption{Maximum achievable ranges, in AU, with a given PPM order and a specified $\theta_{\max}$. Results found using Gaussian beam model, $R = 1/3$, $T_\mathrm{s} = 256\,\mathrm{ns}$, $n_\mathrm{b} = 1.21\cdot 10^7\,\mathrm{phe/s}$, $P_{\mathrm{av}} = 5\,\mathrm{W}$, $P_\mathrm{out}=5$\%, $\lambda = 1064\,\mathrm{nm}$, Link Margin $3\,\dB$.}
\footnotesize
\begin{tabular}{l|C{1.6cm}C{1.6cm}C{1.6cm}C{1.6cm}C{1.6cm}C{1.6cm}C{1.6cm}}
\toprule
 & \multicolumn{7}{c}{\textbf{PPM Order / Peak Laser Power [W]}} \\
 $\theta_{\max} [\mathrm{\upmu rad}]$ & 256 / 1600 & 128 / 800 & 64 / 400 & 32 / 200 & 16 / 100 & 8 / 50 & 4 / 25 \\
\midrule
1.00 & 0.539 & 0.399 & 0.295 & 0.219 & 0.162 & 0.120 & 0.089\\
0.50 & 2.155 & 1.596 & 1.182 & 0.875 & 0.647 & 0.480 & 0.355\\
0.35 & 4.397 & 3.256 & 2.411 & 1.785 & 1.320 & 0.979 & 0.725\\
0.20 & 13.470 & 9.974 & 7.385 & 5.468 & 4.044 & 2.998 & 2.220\\
0.15 & 23.807 & 17.628 & 13.053 & 9.665 & 7.148 & 5.299 & 3.924\\
0.10 & 53.880 & 39.896 & 29.541 & 21.874 & 16.178 & 11.993 & 8.880\\
\midrule
Data Rate [$\mathrm{kbps}$] & 32.33 & 56.58 & 97.00 & 161.66 & 258.66 & 387.99 & 517.32\\
\bottomrule
\end{tabular}
\label{tab:MaxDistTheta}
\end{table*}
\subsection{Maximum Achievable Range}
\label{sec:MaxDist}
The effective gain defined above can be used to analyze the maximum achievable range in optical deep space links in which both antennas are affected by miss-pointing errors. 
A scenario where this analysis is particularly relevant is represented by an optical link between a deep space probe and a second spacecraft orbiting Earth and acting as a relay towards the ground station.

The analysis is carried out for a wavelength $\lambda = 1064\,\mathrm{nm}$. 
We choose the \ac{SCPPM} parameters to minimize the smallest received power such that $P_e \leq P_e^*$, where $P_e$ is the block error probability and $P_e^*$ is its target value.
Accordingly, we set the \ac{PPM} slot time as $T_\mathrm{s} = 256\,\mathrm{ns}$, the convolutional code rate as $R = 1/3$, and the guard time between two adjacent \ac{PPM} symbols, following the CCSDS recommendation, to $\alpha_{\mathrm{gt}} = 25\%$. 
More specifically, this constraint imposes $M/4$ idle slots between any two adjacent \ac{PPM} symbols to guarantee the laser recharge. 
Moreover, we consider an efficiency equal to $-5\,\dB$ for both the transmit and the receive antennas, a supplementary detection and implementation loss equal to $-4\,\dB$, a mean value of the background noise flux $n_\mathrm{b} = 1.21 \cdot 10^7\,\mathrm{phe/ns}$, and $P_e^*= 10^{-4}$ \cite{BisHemPia:10}. 
Concerning pointing losses, we take $P_\mathrm{out} = 5\%$ when applying the probabilistic approach and, for both the deterministic approach and the probabilistic one, we adopt the Gaussian beam model (worst case). 
Furthermore, we assume that the transmit and receive antennas are equal to each other in terms of both gain and pointing accuracy parameter $\sigma_\theta$ or $\theta_\mathrm{max}$. 
Applying the analysis described in Section~\ref{sec:Optimization}, we design the antenna gains to achieve the maximum effective gain. 
Lastly, in compliance with \ac{ITU} recommendations \cite{ITU:06} the link margin is set to $3\,\dB$ and the average transmit power to $5\,\mathrm{W}$. 
No atmospheric attenuation is considered since, as previously mentioned, the receiver is assumed to be a relay orbiting outside the Earth atmosphere.

Table~\ref{tab:MaxDistSigma} and Table~\ref{tab:MaxDistTheta} show the maximum achievable ranges in \ac{AU} (where $1\,\mathrm{AU} = 149,597,871\,\mathrm{Km}$) for both pointing loss approaches. 
In both tables, each column refers to one specific \ac{PPM} order $M$ (the number of slots in one \ac{PPM} symbol), which influences the minimum received power such that $P_e \leq P_e^*$, the transmit peak power, and the information data rate. 
On the other hand, each row refers to a specific value of the pointing accuracy parameter.

The obtained results reveal how the pointing accuracy is a very critical parameter in the optical link design. 
As $\sigma_\theta$ or $\theta_{\max}$ varies, we observe large achievable rate variations for the same \ac{PPM} order, target performance, and information data rate. 
This is related to the impossibility to achieve unbounded gains when pointing losses are included in the model, especially if they are included both in the transmitter and in the receiver. 
We can also compare the results one would obtain without including pointing losses in the analysis, to remark the importance of a correct evaluation pointing losses in deep space inter satellite optical links. 
Considering, for example, the case $M = 256$, using the deterministic method with $\theta_{\max} = 0.35$ \si{\micro}\si{\radian}, using the same antenna gains but neglecting pointing losses, the maximum achievable range would be $11.63\,\mathrm{AU}$, almost three times larger than the value $4.397\,\mathrm{AU}$ reported in Table~\ref{tab:MaxDistTheta}.

\begin{table}[t]
\centering
\caption{Example of Mars \ac{TM} optical link budget with pointing losses on both the transmitter and the receiver. Maximum effective system gain.}
\scriptsize
\begin{tabular}{lC{1cm}C{1cm}C{1cm}}
\toprule
\textbf{Link Parameter} & dB &  & Units \\
\midrule
\rowcolor[gray]{.95} 
\textit{Signaling and Fixed Parameter} & & & \\
PPM Order & & 64 & \\
Convolutional Code Rate & & 1/3 & \\
Slot Time & & 256 & ns\\
Guard Time & & 25 & \% \\
Mean Noise Flux & -19.17 & 1.21e-02 & phe/ns\\
Mean Noise Flux per slot & & 3.10 & phe/slot \\
$\theta_{\max}$, Gaussian Beam & & 0.35 & \si{\micro}\si{\radian} \\
\midrule
\rowcolor[gray]{.95}
\textit{Laser Transmitter} & & & \\
Average Laser Power & 6.99 & 5.00 & W \\
Peak Laser Power & 26.02 & 400 & W \\
Wavelength & & 1064 & nm \\
\midrule
\rowcolor[gray]{.95}
\textit{Deep Space Orbiter} & & & \\
Far-Field Antenna Gain & 129.00 &  &  \\
Transmitter Efficiency & -5.00 & & \\
\midrule
\rowcolor[gray]{.95}
\textit{Range} & & & \\
Space Loss & -373.49 & 2.68 & AU  \\
\midrule
\rowcolor[gray]{.95}
\textit{Near Earth Orbiter} & & &\\
Receiver Gain & 129.00 &  & \\
Receiver Efficiency & -5.00 & & \\
\midrule
\rowcolor[gray]{.95}
\textit{Other} & & &\\
Detection/Implementation Losses & -4.00 & & \\
Pointing Loss & -8.45 &  & \\
\midrule
\rowcolor[gray]{.95}
\textit{Link Performance} & & &\\
Average Received Power & -130.96 & & W  \\
Average Received Photon Flux & -33.67 & 4.30e-04 & phe/ns \\
Minimum Average Received Power & -133.05 &  & W  \\
Minimum Average Received Photon Flux & -35.76 & 2.65e-04 & phe/ns \\
Link Margin & 2.09 & & \\
FER target & & 9.00e-05 & \\
Information Data Rate & & 0.10 & Mbps\\
\bottomrule
\end{tabular}
\label{tab:LinkBudgetMars}
\end{table}

\subsection{Link Budget}
Finally we provide an example of \ac{TM} optical link budget. 
In particular, we consider a deep space probe orbiting Mars and communicating with a relay orbiting the Earth. 
There are three fundamental steps in the link budget procedure. 
Firstly, the optical gain and the pointing losses are derived, through the procedure described in Sec.~\ref{sec:Optimization}, from the pointing requirements and the wavelength. 
Hereafter we use the deterministic approach with $\theta_\mathrm{\max} = 35$ \si{\micro}\si{\radian} and we assume the same gain and the same pointing accuracy parameter for both the transmit and receive antennas. 
This results in an optimal gain of $129\,\dB$ when the Gaussian beam is considered. 
Next, through the optical link equation we compute the received average number of signal photons per second, $n_\mathrm{s}$, as \cite{BisHemPia:10, BisPia:03, HemBisDjo:11, LyrPan:19}
\begin{align}
\label{eq:ns}
    n_\mathrm{s} = P_\mathrm{av}\, G_\rmt\, \eta_\rmt\, \left( \frac{\lambda}{4\pi r} \right)^2\, G_\rmr\, \eta_\rmr\, L_\mathrm{p,tot}\, L_\mathrm{other}\, \frac{\lambda}{h\,c}
\end{align}
where $G_t$ and $G_r$ are the transmit and receive antenna gains, $\eta_t$ and $\eta_r$ are their corresponding efficiencies, $\lambda$ is the laser wavelength, $h$ and $c$ are the Planck's constant and the speed of light, $r$ is the distance between the antennas, $P_\mathrm{av}$ is the average laser power, $L_\mathrm{p,tot}$ are the pointing losses, and $L_\mathrm{other}$ are extra losses (detection and implementation ones).  
Finally, we find a suitable configuration of the \ac{SCPPM} parameters $M$, $R$, and $T_\mathrm{s}$. 
The goal is to have
\begin{align}
    \mathrm{LM}\,[\dB] = n_\mathrm{s}\,[\dB] - n_\mathrm{s}^{\min}\,[\dB]\,.
\end{align}
where $\mathrm{LM}\,[\dB]$ is the requested link margin and $n_\mathrm{s}^{\min}=n_\mathrm{s}^{\min}(M,R,T_\mathrm{s},n_\mathrm{b})$ is the value of $n_{\mathrm{s}}$ such that $P_e=P_e^*$. The information data rate is then given by
\begin{align}
        B_\mathrm{r} = \frac{ 15120\, R  - 34}{M T_\mathrm{s} \left( 15120/ \log_\mathrm{2}M \right) \alpha_{\mathrm{gt}}}
        \label{eq:DataRate}
\end{align}
where $\alpha_\mathrm{gt}$ is the guard time, $15120$ is the \ac{SCPPM} codeword length, and $34$ is the number of \ac{CRC} and termination bits.

Table~\ref{tab:LinkBudgetMars} summarizes the parameters and the resulting data. The maximum Mars-Earth distance is considered, to provide a worst case analysis.
From \eqref{eq:ns}, $n_\mathrm{s} = -33.67\,\mathrm{dB\, phe/ns}$ is obtained. 
In order to guarantee a link margin compliant with the \ac{ITU} recommendation \cite{ITU:06}, a PPM order $M = 64$, a convolutional code rate $R = 1/3$, and a $T_\mathrm{s} = 256\,\mathrm{ns}$ have been selected, leading to $n_\mathrm{s}^{\mathrm{min}} = -35.76\,\dB\, \mathrm{phe/s}$, information data rate of $100\,\mathrm{kbps}$ and a link margin of $2.09\,\dB$. 
Again, no atmospheric losses have been considered as the receiver is assumed outside the Earth atmosphere.
A similar analysis was carried out assuming a deep space spacecraft orbiting Venus. Owing to the smaller worst-case distance, equal to $1.74$ AU, in the Venus case it is possible to choose a \ac{PPM} order $M = 64$, a code rate $R = 1/3$, and a slot time $T_\mathrm{s} = 64 \, \mathrm{ns}$ to achieve a data rate of $390\,\mathrm{kbps}$, with a link margin of $2.17\,\mathrm{dB}$.

Considering no pointing losses in the link budget and using the same input data, the modulation parameters would be mistakenly chosen as $M = 64$, $R = 1/3$, and $T_\mathrm{s} = 16\,\mathrm{ns}$ leading, in the Mars case, to $B_{\mathrm{r}}=1.55\,\mathrm{Mbps}$ with an expected link margin of $3.07\,\dB$. However, due to miss-pointing, the actual link margin would be $-5.38\,\dB$. This highlights the importance of an accurate miss-pointing errors estimation and the fundamental role played by antenna gain optimization.

\section{Conclusion}\label{sec:conclusions}

In this paper, we investigated the effect of pointing losses on optical links in a general scenario where both the transmit and receive antennas are affected by mechanical noise. 
We defined two approaches which differ in terms of knowledge of the angular error statistics. 
As opposed to the deterministic approach, the probabilistic one formulates the problem in terms of an outage probability. 
The developed framework allows evaluating the pointing losses even in presence of different error distributions and radiation patterns, for example, due to an asymmetric design of the transmitter and the receiver.
Results have been presented in terms of effective system gain optimization, maximum achievable ranges in deep space applications, and inter-planetary optical link budgets.


\vspace{-0.4mm}
\section*{Acknowledgment}
This work was supported in part by \ac{ESA}/ESTEC under contract 4000132053/20/NL/FE.


\end{document}